\documentclass{aastex631}

\begin{document}

\title{Estimating the Absolute Parameters of W UMa-type Binary Stars Using Gaia DR3 Parallax}

\author{Atila Poro}
\altaffiliation{poroatila@gmail.com}
\affiliation{Astronomy Department of the Raderon AI Lab., BC., Burnaby, Canada}

\author{Mahya Hedayatjoo}
\affiliation{Department of Physics, Iran University of Science and Technology, Tehran, Iran}

\author{Maryam Nastaran}
\affiliation{Faculty of Geology, University of Tehran, Tehran, Iran}

\author{Mahshid Nourmohammad}
\affiliation{Independent Astrophysics Researcher, Tehran, Iran}

\author{Hossein Azarara}
\affiliation{Faculty of Physics, Shahid Bahonar University of Kerman, P.O.Box 76175, Kerman, Iran}

\author{Sepideh AlipourSoudmand}
\affiliation{Independent Astrophysics Researcher, Tehran, Iran}

\author{Fatemeh AzarinBarzandig}
\affiliation{Independent Astrophysics Researcher, Tehran, Iran}

\author{Razieh Aliakbari}
\affiliation{Physics Society of Iran (PSI), Tehran, Iran}

\author{Sadegh Nasirian}
\affiliation{Faculty of Basic Sciences, University of Guilan, Rasht, Iran}

\author{Nazanin Kahali Poor}
\affiliation{Independent Astrophysics Researcher, Tehran, Iran}

\begin{abstract}
The accuracy of absolute parameters' estimation in contact binary systems is important for investigating their evolution and solving some challenges. The Gaia DR3 parallax is one of the methods used for estimating the absolute parameters, in cases where photometric data is the only one that is available. The use of this method includes advantages and limitations that we have described and examined in this study. We selected 48 contact binary systems whose mass ratios were mostly obtained by spectroscopic data, in addition to a number of photometric studies. The target systems were suitable for $A_V$ and the Re-normalised Unit Weight Error (RUWE), and their absolute parameters were calculated based on Gaia DR3 parallax, observational information, orbital period, and light cure solution from the literature and catalogs. The outcomes of OO Aql differed significantly from those reported in the literature. Upon analyzing the system's light curve with TESS data, we concluded that the stars' temperatures were the reason for this difference, and utilizing Gaia DR3 parallax provided reasonable results. We displayed the target systems on the Hertzsprung-Russell (HR), $q-L_{ratio}$, $P-M_V$, and $logM_{tot}-logJ_0$ diagrams, and the systems are in good agreement with the theoretical fits. We showed that the estimation of absolute parameters with this method might be acceptable if $\Delta a(R_{\odot})$ is less than $\sim0.1(R_{\odot})$. There are open questions regarding the existence of $l_3$ in the light curve analysis and its effect on the estimation of absolute parameters with this method.
\end{abstract}


\section{Introduction}
W UMa-type contact binary systems are eclipsing binaries with a short orbital period, commonly less than a day. They are made up of two components that share an envelope and overfill the inner Roche lobes, so it is possible to have the exchange of mass and energy through the common envelope (\citealt{2005ApJ...629.1055Y}). The late spectral (A-K) type stars appear in the W UMa systems.
They show continuous variations in brightness and equal or almost the same depths in the two eclipse minima, suggesting close temperatures for the two components.

Studies have been done with different results regarding many general features of these types of stellar binary systems. However, investigations continue on contact systems' lower and higher limits for the orbital period, maximum stars' temperatures, maximum temperature differences between two components, mass ratio ($q$), etc. For example, the results of \cite{2021ApJS..254...10L}'s investigation indicate that a binary system cannot be classified as a W UMa-type if it has an orbital period greater than 0.5 days and an effective temperature of more than 7000 K.

The mass and temperature of the companions are on account for the division of contact binary systems into two A and W subtypes (\citealt{1970VA.....12..217B}). Therefore, in A-type contact systems, the more massive component has a higher effective temperature, whereas in W-type contact systems, the less massive component has a higher effective temperature. As a result, a suitable light curve solution and an accurate mass estimate are required to identify the subtype of each system.

As mentioned in many studies, W UMa-type systems are significant astrophysical tools for studying the formation, evolution, and structure of stars (\citealt{1994ASPC...56..228B}, \citealt{2003MNRAS.342.1260Q}, \citealt{2005ApJ...629.1055Y}, \citealt{2007ApJ...662..596L}, \citealt{2008MNRAS.386.1756E}, \citealt{2012JASS...29..145E}). It is possible to get insight into the evolution of stars by studying contact binary systems, including relationships between parameters and important stellar elements like mass, luminosity, and surface gravity. Deepest investigations are required to solve the numerous unanswered concerns and unsolved challenges regarding these systems in our current understanding.

To obtain reliable results and overcome some challenges, it is necessary to measure accurately and use appropriate methods for computations and analysis. One of the issues is estimating absolute parameters after observation and data reduction processes, analysis of orbital period variations, and light curve solution. When just photometric data is available, there are several methods based on various parameter relationship equations that were presented through investigations. The use of Gaia DR3\footnote{\url{https://gea.esac.esa.int/archive/}} parallax is one method that has been utilized recently to estimate the absolute parameters of contact binary systems (\citealt{2019RAA....19...14K}, \citealt{2021AJ....162...13L}, \citealt{2022MNRAS.510.5315P}). Gaia parallax is also used to estimate absolute parameters in other types of variable stars such as Delta Scuti (\citealt{poro2024period}).

In this study, we attempted to investigate different aspects of using this method to estimate parameters. Also, we presented an estimation of absolute parameters for 48 contact binary systems using Gaia DR3 parallax.

\vspace{1.5cm}
\section{Estimation Absolute Parameters}
There are several methods used in studies to estimate absolute parameters when just photometric data are available. For example, the provided equations between orbital period and mass ($P-M$) are used, which are always updated with new studies. \cite{qian2003overcontact} has provided $P-M_1$ equations based on the amount of mass and orbital period of the systems; Similar equations for all subtypes of contact binary systems were also presented by \cite{2021ApJS..254...10L} and \cite{2022MNRAS.510.5315P}. The challenge of using these equations in some contact systems is the selection of the primary component; However, in the \cite{2022MNRAS.510.5315P} study, $M_1$ is related to a more massive star. After determining $M_1$, the mass of the other component can then be calculated using the mass ratio found in the light curve solution. Following that, the system's total mass and orbital period are used to compute semi-major axis $a(R_{\odot})$ using Kepler's third law. The radius of the stars is then determined using the relationship between $a(R_{\odot})$ and $r_{mean}$. Knowing the temperature and radius of each star allows for the calculation of the star's luminosity.

On the other hand, the equation between $a(R_{\odot})$ and the orbital period is used in some studies for the absolute parameter estimation of contact binaries. $P-a$ equations that are appropriate for both A and W subtypes of contact binary systems have been presented by the \cite{2008MNRAS.390.1577G} and \cite{2024PASP..136b4201P} studies. Kepler's third law equation is possibly utilized to estimate the mass of each component with $a(R_{\odot})$, orbital period, and mass ratio.

The use of Gaia parallax has been one of the methods employed by studies in recent years considering Gaia was able to get accurate parallax of millions of stars. Thus, it is possible to determine the absolute magnitude ($M_V$) of the system by utilizing the system's distance ($d$) computed by Gaia's parallax, its maximum apparent magnitude ($V_max$), and extinction coefficient $A_V$. 
The values of $M_{V1}$ and $M_{V2}$ can be calculated using $l_{1,2}/l_{tot}$ from the light curve solution. The bolometric correction (BC) for each component is needed to calculate the bolometric magnitude ($M_{bol1,2}$). It is possible to compute the stars' luminosity ($L_{1,2}$) by employing the well-known $M_{bol}$ and $L$ relationship. The stars' radius ($R$) is then computed using the luminosity, effective temperature ($T_{1,2}$), and Stefan-Boltzmann constant ($\sigma$). It was possible to estimate $a(R_{\odot})$, $M_1$, and $M_2$ using $r_{mean1,2}$, orbital period, and mass ratio. Also, the surface gravity values ($g_{1,2}$) by a well-known equation and the orbital angular momentum ($J_0$) of the system using the \cite{2006MNRAS.373.1483E} study can be calculated. The described equations are as follows:

\begin{equation}\label{eq1}
M_{V(system)}=(V_{max})-5log(d)+5-(A_V)
\end{equation}

\begin{equation}\label{eq2}
M_{V(1,2)}-M_{V(tot)}=-2.5log(\frac{l_{(1,2)}}{l_{(tot)}})
\end{equation}

\begin{equation}\label{eq3}
M_{bol}=M_{V}+BC
\end{equation}

\begin{equation}\label{eq4}
M_{bol}-M_{bol_{\odot}}=-2.5log(\frac{L}{L_{\odot}})
\end{equation}

\begin{equation}\label{eq5}
R=(\frac{L}{4\pi \sigma T^{4}})^{1/2}
\end{equation}

\begin{equation}\label{eq6}
a=\frac{R}{r}
\end{equation}

\begin{equation}\label{eq7}
\frac{a^3}{G(M_1+M_2)}=\frac{P^2}{4\pi^2}
\end{equation}

\begin{equation}\label{eq8}
g=G_{\odot}(\frac{M}{R^2})
\end{equation}

\begin{equation}\label{eq9}
J_0=\frac{q}{(1+q)^2} \sqrt[3] {\frac{G^2}{2\pi}M^5P}
\end{equation}

\vspace{1.5cm}
\section{Examine the Method}
\subsection{Important Parameters}
Is absolute parameter estimation using Gaia DR3 parallax suitable for all contact binary systems? To answer this question, we should discuss the parameters that have a greater impact on the process of estimating absolute parameters and knowing the limitations:

- $V_{max}$(mag.): This is an observed parameter that will play an important role in the accuracy of calculations. Even the accuracy of the $V_{max}$ error affects the reasonableness of absolute parameters' errors. It is important to consider that scattering in the ground-based observations light curve's maxima might result from poor observational conditions or physical causes. Therefore, in cases where there is an unusual dispersion in the maxima, it is better not to use the Gaia parallax process for absolute parameter calculations. Additionally, some catalogs have reported the $V_{max}$ parameter. However, in the absence of observations and no reports by catalogs, their declared apparent magnitude value can be used if the $V_{max}$ error is about half of the system's light variations.

- $d$(pc): A star or stellar system with a reliable parallax value yields a reliable distance. The discussed method of estimating absolute parameters is based on the accuracy of Gaia's parallax. The Gaia project has always been improving and fixing defects, and in 2020, it was able to improve the accuracy of parallaxes by about $30\%$ by presenting the Gaia DR3 catalog (\citealt{2021AA...649A...1G}). Even so, investigations on the systematic zero-point offset of the Gaia DR3 parallaxes suggest that correction is required to the parallax values to make them less biased and more accurate (\citealt{lindegren2021gaia}, \citealt{ren2021gaia}). It seems that the value of this suggested parallax modification lies entirely in the parameter error range, particularly in the presence of $V_{max}$, thus it could be ignored (\citealt{2022MNRAS.510.5315P}).

- $A_v$: Binary systems and stars with galactic coordinates (b) in the range of +5 to -5 indicate high values of $A_V$ and parallax errors (\citealt{2024PASP..136b4201P}). It is difficult to identify the exact value of $A_V$ that is appropriate for estimating absolute parameters using Gaia parallax; however, values more than $\sim0.4$ seem to cause challenges. It is suggested to consider the RUWE from Gaia DR3 for better evaluation of all target systems. According to \cite{lindegren2018re}, the maximum range for the RUWE index should be 1.4.

- $l_{1,2}/l_{tot}$: The importance of this parameter, which is obtained from the light curve solution, is in Equation \ref{eq2} to determine $M_{V1}$ and $M_{V2}$. Actually, the absolute magnitude value of the system is calculated at first, and the $l_{1,2}/l_{tot}$ parameter determines the absolute magnitude value of each companion. Thus, in the case of an invalid light curve analysis, the absolute parameters for each companion are estimated incorrectly and identified in the computation of $a_{1,2}(R_{\odot})$. In the process of estimating absolute parameters using Gaia DR3 parallax, the values of $a_1(R_{\odot})$ and $a_2(R_{\odot})$ are obtained, which should be theoretically the same. However, in calculations, their values should be obtained close to each other, and $a(R_{\odot})$ is calculated from their average. It should be noted that in the studies that use this method to estimate the absolute parameters, there was no discussion of the $l_3$ effect in the process.
\\
\\
\subsection{Sample}
We considered 48 contact binary systems as a sample from literature studies, and most of them have used the results of spectroscopic analysis for mass ratio. The purpose of recalculating the absolute parameters of this number of systems is to investigate some ambiguities in the method using Gaia's parallax. Also, some of these 48 systems have calculated absolute parameters with various other methods, and some have not provided these parameters. It should be noted that the number of investigated systems was much higher, but considering some of the limitations mentioned, such as $A_V$, RUWE, and having all the necessary parameters, their number was reduced.
Therefore, the selected systems had the following parameters, which were required to compute the absolute parameters in this process: $V_{max}$, orbital period, the effective temperature of components, mass ratio, $l_{1,2}/l_{tot}$ in the $V$ band, and $r_{mean1,2}$.

The sample consists of contact binary systems from both the northern and southern hemispheres of the sky, with an orbital period range of 0.256 to 0.622, a maximum apparent magnitude of 7.27 to 16.40, a star1 effective temperature of 4410 K to 7000 K, and a star2 effective temperature of 4341 to 7048 K. Both the A and W subtypes are present in the sample, and there are also some light curve solutions with a third body ($l_3$).
Table \ref{tab1} contains the general information of the target systems, including coordinates from the Simbad\footnote{\url{https://simbad.u-strasbg.fr/simbad/}} database, distance and RUWE are from Gaia DR3, $V_{max}$ and the orbital period from the VSX\footnote{\url{https://www.aavso.org/vsx/index.php}} database.
The last column of Table \ref{tab1} is the calculation results of $A_V$ from the 3D dust map based on the Gaia (\citealt{2019ApJ...887...93G}).
Table \ref{tab2} is a list of references for selected studies between the years 2001 and 2021. Also, Tables \ref{tab3} and \ref{tab4} show the light curve analysis and the absolute parameters from the selected studies, respectively. As shown in Table \ref{tab4}, some studies did not calculate absolute parameters; And the results in this table are the results of calculations with methods other than the use of Gaia parallax by studies. We estimated the absolute parameters of 48 contact systems using Gaia DR3's parallax, whose results for the $L_{1,2}$, $R_{1,2}$, $M_{1,2}$, and $a(R_{\odot})$ parameters are presented in Table \ref{tab5}. Table \ref{tab6} also lists the results of our calculations of some other absolute parameters and the orbital angular momentum of the systems.
\\
\\
\subsection{Discussion}
\begin{figure*}
\begin{center}
\includegraphics[width=\textwidth]{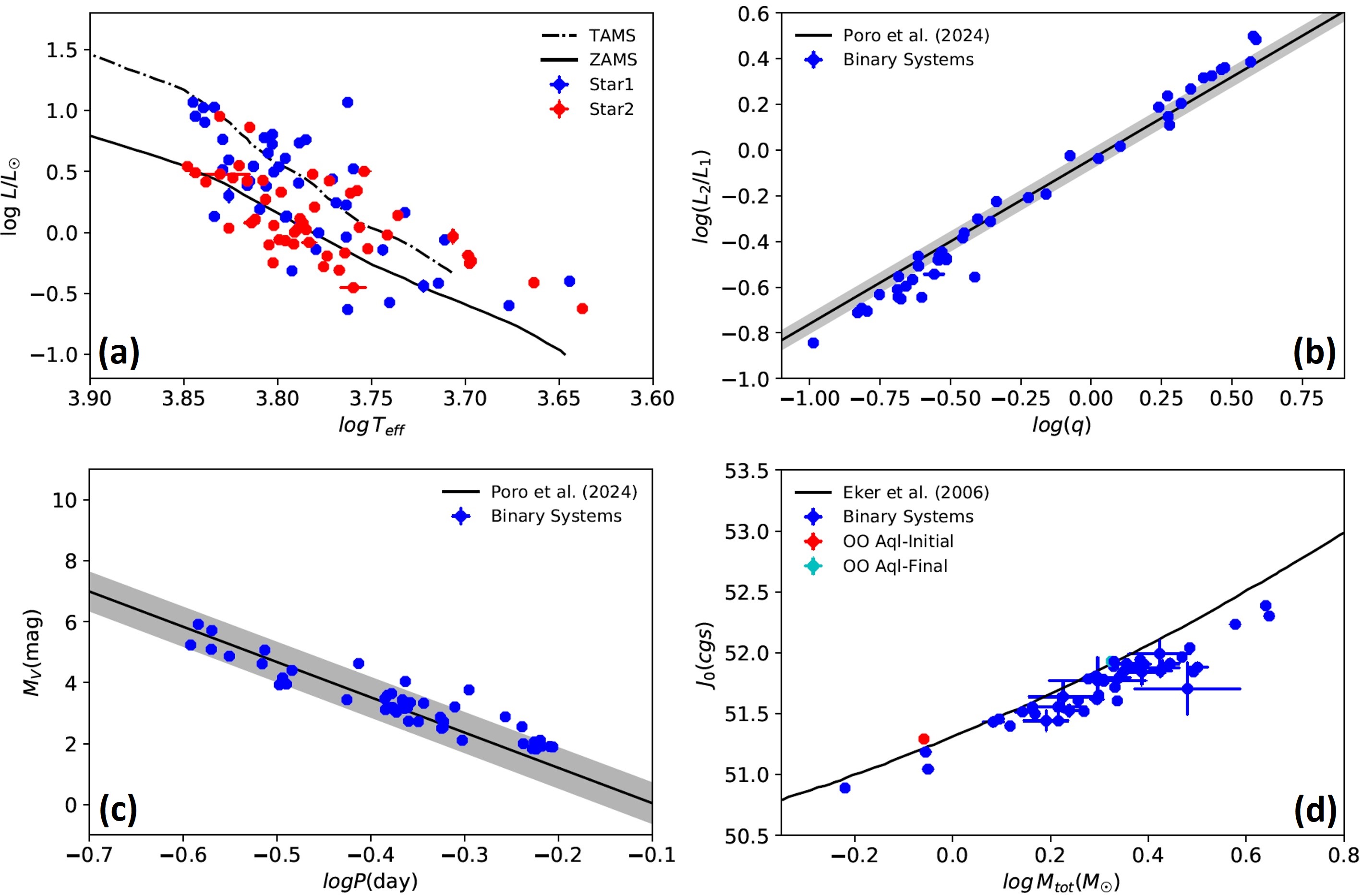}
\caption{Position of the contact binary systems on the HR, $q-L_{ratio}$, $P-M_V$, and $logM_{tot}-logJ_0$ diagrams. Diagram (d) for OO Aql shows its position with absolute parameter estimation based on two different sources: the literature's light curve analysis and the findings in this investigation.}
\label{Fig1}
\end{center}
\end{figure*}

According to the estimation of absolute parameters using Gaia DR3 parallax for 48 contact binary systems and considering the limitations of this method, the results are discussed below:

1. Figures \ref{Fig1}a, \ref{Fig1}b, and \ref{Fig1}c show where the target systems are located on the HR, $q-L_{ratio}$, and $P-M_V$ diagrams. These positions are according to our calculations using the results of literature light curve analysis and Gaia DR3 parallax (Tables \ref{tab5} and \ref{tab6}). The HR diagram includes the Terminal-Age Main Sequence (TAMS) and the Zero-Age Main Sequence (ZAMS) lines. The linear fit in diagrams $q-L_{ratio}$ and $P-M_V$ are from the \cite{2024RAA....24a5002P} study. The position of the systems is acceptable according to the theoretical fits in Figures \ref{Fig1}b, and \ref{Fig1}c.

2. The $logM_{tot}-logJ_0$ diagram (Figure \ref{Fig1}d) shows that the target systems, except one of them, are placed in a contact binary systems region. Figure \ref{Fig1}d shows our estimation using the Gaia DR3 parallax (Tables \ref{tab5} and \ref{tab6}).

As can be seen in the $logM_{tot}-logJ_0$ diagram, the OO Aql system is located above the parabolic boundary from the \cite{2006MNRAS.373.1483E} study, and in the detached binaries region. We used the light curve analysis results of the \cite{2016RAA....16....2L} study, and they used the mass ratio from the \cite{2007AJ....133.1977P} study that reported $q$ by spectroscopic observations.
We tried to understand when the results of our estimations showed that the mass of the stars of OO Aql is less than $0.5M_{\odot}$, while in both studies \cite{2007AJ....133.1977P} and \cite{2016RAA....16....2L}, they declared the mass to be about Sun-like.
As it is clear in Equation \ref{eq9}, masses and orbital period parameters are effective in the calculation of $J_0$. The orbital period of this system is similar in different catalogs and did not change greatly during the studies. After a closer look based on the information we have today thanks to space telescopes, we found that the effective temperatures of the stars in the light curve analysis are very different from what TESS and Gaia DR2 reported.
\cite{2016RAA....16....2L} determined 6100 K for the primary star, and in their observational light curve, the primary minimum is a deeper one. TESS and Gaia DR2 reported $5409\pm262$ K and 5168 K for this system. We did not have their ground-based observational data from 2013, so we decided to re-analyze the light curve using TESS data to clarify.

We used TESS data (HLSP mission) from sector 54 with an exposure time of 600 seconds. OO Aql (TIC 69595756) observations began on July 9, 2022, by TESS. The light curve of the OO Aql system was analyzed using the PHOEBE 2.4.9 version (\citealt{2016ApJS..227...29P}, \citealt{2020ApJS..250...34C}). The bolometric albedo and gravity-darkening coefficients were assumed to have values of $A_1=A_2=0.5$ \cite{1969AcA....19..245R} and $g_1=g_2=0.32$ \cite{1967ZA.....65...89L}. The limb darkening coefficients came from the PHOEBE tables, and we modeled the stellar atmosphere using the \cite{castelli2004new} study. We set the initial temperature from the TESS reported for the primary star. The optimization tool in the PHOEBE code was used to improve the output of the light curve solution. Additionally, the asymmetry in the light curve's maxima indicates that a cold starspot on the primary component is required for the light curve solution (\citealt{1951PRCO....2...85O}). Then, we estimated the absolute parameters using Gaia DR3 parallax. As a result, the mass of stars approached the values reported in the \cite{2007AJ....133.1977P} study using spectroscopic data. Also, orbital angular momentum is estimated as $J_0=51.931\pm0.014$.
Table \ref{tab7} presents the light curve analysis results and estimated absolute parameters.

\begin{figure*}
\begin{center}
\includegraphics[width=\textwidth]{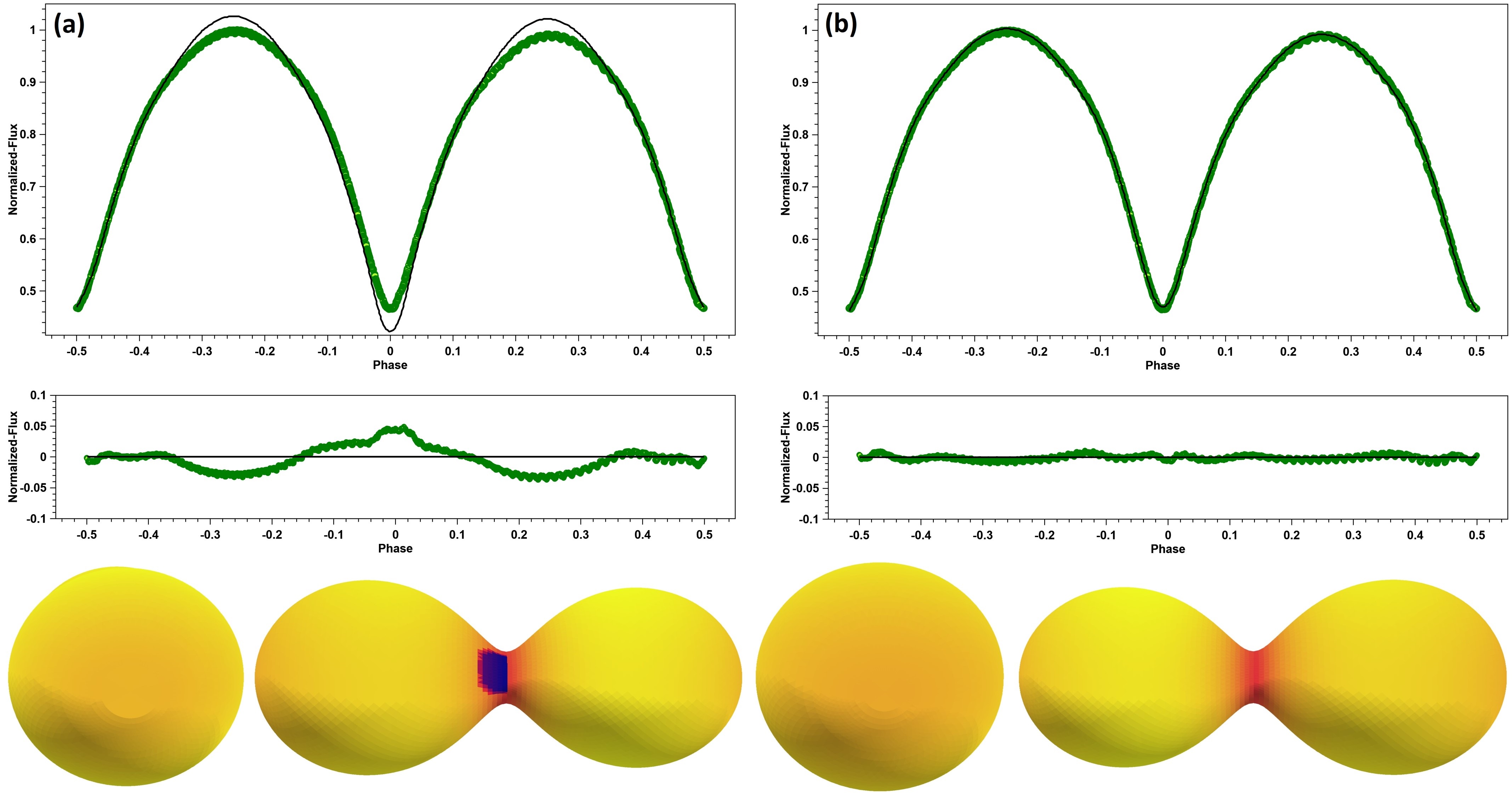}
\caption{The green dots show the OO Aql system's light curve that was observed by TESS, and the synthetic light curves are black curves. (a) is a theoretical light curve from the results of the literature, and (b) is based on the outcome of this study. The geometric structure of the system in phases 0, 0.25, 0.5, and 0.75 is presented.}
\label{Fig2}
\end{center}
\end{figure*}

Figure \ref{Fig2} shows TESS data of OO Aql and synthetic light curves by \cite{2016RAA....16....2L} and this study. The geometric structure of the system with a cold starspot is shown in Figure \ref{Fig2}. As can be seen in Figure \ref{Fig1}d, the location of this system was modified and placed in the area of contact binary systems. OO Aql showed how input results from observations, orbital period variations, and light curve analysis can be important in estimating absolute parameters using Gaia's parallax.

\begin{figure}
\begin{center}
\includegraphics[scale=0.76]{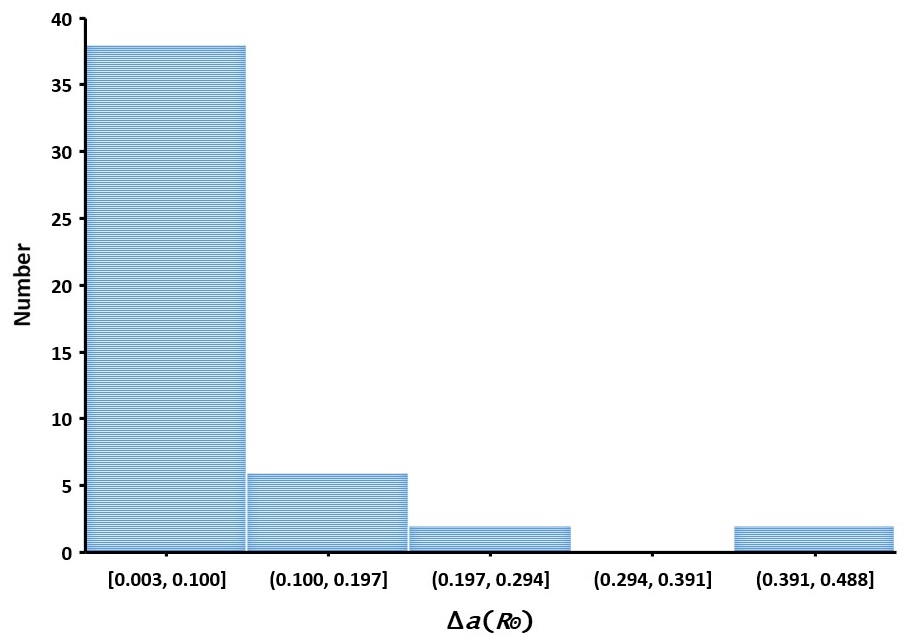}
\caption{The $\Delta a(R_{\odot})$ distribution in the sample.}
\label{Fig3}
\end{center}
\end{figure}

3. Theoretically, we expect $a_1(R_{\odot})$ and $a_2(R_{\odot})$ to be equal. However, most of the time, when we are faced with observational data and their analysis, we can only expect the $\Delta a(R_{\odot})$ to be a small number. According to the systems in our sample and the histogram in Figure \ref{Fig3}, about $79\%$ have a $\Delta a(R_{\odot})$ less than 0.1. Therefore, this range may be considered acceptable based on the sample. The lowest $\Delta a(R_{\odot})$ is related to the studies of systems BO Ari and HH UMa with a value of 0.003. The two with the largest $a(R_{\odot})$ are NR Cam (0.404) and V404 Peg (0.424). Regarding these systems, they cannot be considered careless due to $q$-search since both BO Ari and NR Cam systems only used photometric data; Also, among these systems, there are both total and partial eclipses. Therefore, the cause of low or high values in $\Delta a(R_{\odot})$ can only be considered due to the accuracy of light curve analysis.
However, this is not a definitive parameter for examination, and as stated about the OO Aql system, its $\Delta a(R_{\odot})$ was acceptable. 

4. The OO Aql contact binary system shows that the acquired absolute parameters will be unacceptable if the initial effective temperatures are not selected appropriately. This is considered one of the benefits of using the Gaia parallax method to estimate absolute parameters compared to empirical equations. However, in addition to the impact of effective temperatures, other unsuitable elements could also affect the outcomes. The orbital period, among other characteristics, could have the least effect on the outcomes since, in most cases, the calculations are accurate and correct to three decimal places. However, parameters $V_{max}$ and $A_V$ will have the same effect as the effective temperature because the first parameter, which is $M_V$, is calculated based on them. After them, $l_{1,2}/l_{tot}$ and $r_{mean1,2}$ have more effects that show themselves in $\Delta a(R_{\odot})$ in cases of improper accuracy.

The results of estimating the absolute parameters using Gaia DR3 parallax for systems whose mass ratios were obtained using spectroscopic data show much greater agreement compared to studies that used only photometric analysis. The ASAS J035020-8017.4 system from the \cite{2016NewA...46...94S} study was specifically chosen to highlight the significance of suitable photometric data in addition to input parameter accuracy. The photometric light curve of this system in the \cite{2016NewA...46...94S} study shows a lot of scattering, so the minima are not clear. Therefore, it is unacceptable that the masses of the primary and secondary stars were found to be 7.924 (443), and 1.838 (103), respectively. Unfortunately, the ASAS J035020-8017.4 system did not have another study for comparison. Also, the \cite{2016NewA...46...94S} study's effective temperature results for the stars are lower by about 400 K and 500 K compared to TESS and Gaia DR3's reported temperatures, respectively. Therefore, the results of the light curve analysis of this system are not acceptable due to the data and an inappropriate initial temperature.

5. According to Table \ref{tab3}, eight systems analyzed the light curves with a third body. Moreover, Table \ref{tab6} indicates that for this number of systems, the existence of $l_3$ could affect the values of the absolute parameters but not the increase or decrease of $\Delta a(R_{\odot})$. Many more samples are needed to investigate the effect of $l_3$ in $\Delta a(R_{\odot})$. It should be noted that currently there is no way to apply $l_3$ to calculations of absolute parameters using Gaia parallax.

\vspace{1.5cm}
\section{Summary and Conclusion}
The Gaia DR3 parallax is used to estimate the absolute parameters of contact binary systems due to its appropriate accuracy. In this calculation method, observation parameter ($V_{max}$), orbital period ($P$), light curve solution ($T_1$, $T_2$, $q$, $l_{1,2}/l_{tot}$, $r_{mean1,2}$), and some other calculation parameters ($A_V$, B$C_{1,2}$) are used. Therefore, accuracy is considered in the process from observation to estimation of absolute parameters. This method has been used in recent years, particularly when the studies are limited to using photometric data. Therefore, it is imperative to examine the limits and efficacy of this way.

We selected 48 systems that were suitable in terms of $A_V$ and RUWE and also had Gaia DR3 parallax for this study. In $90\%$ of these studies, $q$ was derived using spectroscopic data or results, and in the remaining studies, only photometric data was available. We utilized the light curve solution from these investigations (Table \ref{tab3}) to determine the absolute parameters using Gaia DR3 Parallax. Some required elements were obtained from the catalogs and then the absolute parameters were estimated using equations \ref{eq1} to \ref{eq9}. Figure \ref{Fig1} shows the results of these calculations. Most of the systems were in good agreement with the theoretical fits in diagrams \ref{Fig1}b,c,d. OO Aql was the name of the system that was placed in the detached binary systems region in the $logM_{tot}-logJ_0$ diagram. We used TESS data and light curve analysis to check this system. It was found that the main cause of the incorrect estimation of the absolute parameters was the stars' temperatures. The new results of the absolute parameters of OO Aql are presented in Table \ref{tab7}; they are in good agreement with the spectroscopic results.

$\Delta a(R_{\odot})$ is one of the crucial factors that indicate the light curve analysis or the parameters utilized for the absolute parameter estimate may not be reliable. According to our results, its appropriate value should be less than 0.1. However, we emphasize that this parameter alone cannot be considered to check the results.

The systems whose absolute parameters were estimated using Gaia DR3 parallax in this study can be examined one by one. We only focused on the results of the two most specific systems in terms of inappropriate results. Estimating absolute parameters using this method is not acceptable if there is scattered data, an incorrect initial temperature, or low-precision light curve analysis.

Considering the accuracy of Gaia DR3 parallax and paying attention to the limitations, this estimation process can have good results, especially when there is only photometric data available. In future studies, it is necessary to consider the limitations mentioned in this study to use this method. One of these limits is $l_3$, and the findings might not be reliable if the light curve solution includes $l_3$.

\vspace{1.5cm}
\section*{Acknowledgements}
This manuscript was prepared by the BSN project (\url{https://bsnp.info/}).
We have made use of data from the European Space Agency (ESA) mission Gaia (\url{http://www.cosmos.esa.int/gaia}), processed by the Gaia Data Processing and Analysis Consortium (DPAC). The SIMBAD database was utilized by this study, which is operated by CDS in Strasbourg, France (\url{http://simbad.u-strasbg.fr/simbad/}). We sincerely thank Ehsan Paki for his scientific and coding assistance. We thank Kai Li for his helpful comments.

\vspace{1.5cm}
\section*{ORCID iDs}
\noindent Atila Poro: 0000-0002-0196-9732\\
Mahya Hedayatjoo: 0000-0002-0192-215X\\
Maryam Nastaran: 0006-9150-3392\\
Mahshid Nourmohammad: 0009-0006-0476-2055\\
Hossein Azarara: 0009-0003-2631-6329\\
Sepideh AlipourSoudmand: 0009-0005-1048-0866\\
Fatemeh AzarinBarzandig: 0009-0004-7836-7686\\
Razieh Aliakbari: 0009-0007-8508-2357\\
Sadegh Nasirian: 0009-0001-8140-1505\\
Nazanin Kahali Poor: 0009-0007-5785-7303\\

\clearpage

\begin{table*}
\caption{General specifications of target systems.}
\centering
\begin{center}
\footnotesize
\begin{tabular}{c | c | c | c | c | c | c | c}
\hline
\hline
System & RA.(J2000) & Dec.(J2000) & $d$(pc) & RUWE & $V_{max}$(mag.) & P(day) & $A_v$\\
\hline
AK Her	&	$	17$ $13$ $57.8236	$	&	$	+16$ $ 21 $ $00.6182	$	&	$	104.62	\pm	0.18	$	&	0.961	&	8.29	&	0.4215228	&$	0.078	\pm	0.001	$	\\
AO Aqr	&	$	22$ $ 11 $ $31.9092	$	&	$	-22$ $ 47 $ $16.5454	$	&	$	356.47	\pm	3.07	$	&	0.926	&	11.04	&	0.4893432	&$	0.074	\pm	0.001	$	\\
AP Leo	&	$	11$ $ 05 $ $05.0213	$	&	$	+05$ $ 09 $ $06.4048	$	&	$	144.81	\pm	0.35	$	&	1.025	&	9.32	&	0.4303580	&$	0.078	\pm	0.001	$	\\
AQ Tuc	&	$	00$ $ 17 $ $21.5138	$	&	$	+71$ $ 54 $ $56.8472	$	&	$	360.18	\pm	1.53	$	&	0.969	&	9.91	&	0.5948387	&$	0.071	\pm	0.001	$	\\
ASAS J035020-8017.4	&	$	03$ $ 50 $ $19.6100	$	&	$	+80$ $ 17 $ $22.6196	$	&	$	923.69	\pm	8.86	$	&	1.073	&	11.94	&	0.6224100	&$	0.221	\pm	0.001	$	\\
ASAS J063546+1928.6	&	$	06$ $ 35 $ $46.2147	$	&	$	+19$ $ 28 $ $28.0424	$	&	$	270.89	\pm	1.52	$	&	1.051	&	9.91	&	0.4755130	&$	0.210	\pm	0.001	$	\\
AU Ser	&	$	15$ $ 56 $ $49.4687	$	&	$	+22$ $ 16 $ $01.5907	$	&	$	162.86	\pm	0.38	$	&	1.069	&	10.80	&	0.3864965	&$	0.110	\pm	0.001	$	\\
BO Ari	&	$	02$ $ 12 $ $08.7727	$	&	$	+27$ $ 08 $ $18.2311	$	&	$	141.58	\pm	0.43	$	&	0.954	&	9.77	&	0.3181950	&$	0.082	\pm	0.001	$	\\
BX Dra	&	$	16$ $ 06 $ $17.3670	$	&	$	+62$ $ 45 $ $46.0898	$	&	$	520.27	\pm	4.65	$	&	0.856	&	10.63	&	0.5790246	&$	0.048	\pm	0.001	$	\\
CU Tau	&	$	03$ $ 47 $ $36.9104	$	&	$	+25$ $ 23 $ $15.8452	$	&	$	351.02	\pm	2.75	$	&	0.975	&	11.50	&	0.4122200	&$	0.298	\pm	0.002	$	\\
DD Com	&	$	12$ $ 28 $ $46.2477	$	&	$	+21$ $ 43 $ $33.3518	$	&	$	675.51	\pm	11.16	$	&	1.032	&	14.30	&	0.2692061	&$	0.054	\pm	0.001	$	\\
DN Boo	&	$	13$ $ 51 $ $42.0430	$	&	$	+14$ $ 18 $ $05.9467	$	&	$	453.14	\pm	4.35	$	&	0.989	&	11.06	&	0.4475640	&$	0.058	\pm	0.001	$	\\
DN Cam	&	$	04$ $ 42 $ $46.2426	$	&	$	+72$ $ 58 $ $41.8759	$	&	$	156.53	\pm	0.50	$	&	0.951	&	8.21	&	0.4983104	&$	0.130	\pm	0.001	$	\\
EF Boo	&	$	14$ $ 32 $ $30.5386	$	&	$	+50$ $ 49 $ $40.6868	$	&	$	160.93	\pm	0.34	$	&	1.066	&	9.23	&	0.4205161	&$	0.021	\pm	0.001	$	\\
EF Dra	&	$	18$ $ 05 $ $30.4690	$	&	$	+69$ $ 45 $ $15.7509	$	&	$	287.28	\pm	1.17	$	&	1.055	&	10.38	&	0.4240310	&$	0.053	\pm	0.001	$	\\
EH Cnc	&	$	08$ $ 26 $ $18.3530	$	&	$	+20$ $ 52 $ $49.7272	$	&	$	495.10	\pm	4.89	$	&	1.079	&	11.73	&	0.4180366	&$	0.085	\pm	0.001	$	\\
EI CVn	&	$	14$ $ 02 $ $05.5579	$	&	$	+34$ $ 02 $ $40.0367	$	&	$	149.36	\pm	0.39	$	&	1.357	&	11.82	&	0.2607661	&$	0.030	\pm	0.001	$	\\
EQ Cep	&	$	00$ $ 47 $ $33.3762	$	&	$	+85$ $ 16 $ $24.2542	$	&	$	1650.12	\pm	95.78	$	&	0.969	&	16.40	&	0.3069522	&$	0.242	\pm	0.001	$	\\
GSC 3553-845	&	$	18$ $ 57 $ $15.3479	$	&	$	+51$ $ 16 $ $31.6251	$	&	$	716.24	\pm	7.12	$	&	1.258	&	12.56	&	0.4354670	&$	0.106	\pm	0.001	$	\\
GSC 804-118	&	$	08$ $ 31 $ $25.2564	$	&	$	+11$ $ 48 $ $13.0960	$	&	$	817.58	\pm	14.43	$	&	1.018	&	13.60	&	0.3236934	&$	0.085	\pm	0.001	$	\\
GSC 2936-478	&	$	06$ $ 35 $ $40.5350	$	&	$	+42$ $ 04 $ $14.8356	$	&	$	873.60	\pm	11.18	$	&	1.092	&	13.40	&	0.4386900	&$	0.346	\pm	0.002	$	\\
GW Cnc	&	$	08$ $ 48 $ $12.6908	$	&	$	+21$ $ 07 $ $13.7982	$	&	$	328.26	\pm	1.86	$	&	0.949	&	12.52	&	0.2814130	&$	0.068	\pm	0.001	$	\\
GZ And	&	$	02$ $ 12 $ $14.0689	$	&	$	+44$ $ 39 $ $36.2081	$	&	$	168.48	\pm	0.53	$	&	1.024	&	10.83	&	0.3050169	&$	0.074	\pm	0.001	$	\\
HH UMa	&	$	11$ $ 04 $ $48.1520	$	&	$	+35$ $ 36 $ $26.6165	$	&	$	261.57	\pm	1.09	$	&	1.050	&	10.58	&	0.3754920	&$	0.051	\pm	0.001	$	\\
HI Dra	&	$	18$ $ 33 $ $24.3606	$	&	$	+58$ $ 42 $ $23.3479	$	&	$	264.53	\pm	0.96	$	&	1.208	&	9.02	&	0.5974190	&$	0.085	\pm	0.001	$	\\
HI Pup	&	$	07$ $ 33 $ $38.2096	$	&	$	-50$ $ 07 $ $25.0452	$	&	$	250.11	\pm	0.67	$	&	0.981	&	10.40	&	0.4326180	&$	0.250	\pm	0.001	$	\\
KIC 3221207	&	$	19$ $ 09 $ $21.8128	$	&	$	+38$ $ 19 $ $32.4282	$	&	$	670.65	\pm	4.28	$	&	1.006	&	11.97	&	0.4738220	&$	0.335	\pm	0.001	$	\\
MQ UMa	&	$	11$ $ 21 $ $41.0631	$	&	$	+43$ $ 36 $ $52.7124	$	&	$	573.79	\pm	7.27	$	&	0.989	&	11.57	&	0.4760580	&$	0.058	\pm	0.001	$	\\
NO Cam	&	$	04$ $ 14 $ $51.4457	$	&	$	+75$ $ 20 $ $40.7481	$	&	$	647.32	\pm	4.74	$	&	1.061	&	12.67	&	0.4307812	&$	0.365	\pm	0.001	$	\\
NR Cam	&	$	07$ $ 54 $ $30.5738	$	&	$	+78$ $ 06 $ $45.3384	$	&	$	124.91	\pm	0.18	$	&	0.960	&	10.76	&	0.2558850	&$	0.041	\pm	0.001	$	\\
OO Aql	&	$	19$ $ 48 $ $12.6526	$	&	$	+09$ $ 18 $ $32.3750	$	&	$	117.27	\pm	0.23	$	&	1.018	&	9.20	&	0.5067920	&$	0.089	\pm	0.001	$	\\
RR Cen	&	$	14$ $ 16 $ $57.2209	$	&	$	-57$ $ 51 $ $15.6569	$	&	$	106.96	\pm	0.41	$	&	0.992	&	7.27	&	0.6056903	&$	0.211	\pm	0.001	$	\\
RV CVn	&	$	13$ $ 40 $ $18.1620	$	&	$	+28$ $ 18 $ $21.5427	$	&	$	415.24	\pm	3.28	$	&	0.915	&	13.84	&	0.2695662	&$	0.033	\pm	0.001	$	\\
RV Psc	&	$	01$ $ 19 $ $41.0893	$	&	$	+31$ $ 12 $ $04.9786	$	&	$	452.64	\pm	4.59	$	&	1.188	&	11.30	&	0.5539915	&$	0.139	\pm	0.001	$	\\
SS Com	&	$	12$ $ 49 $ $39.0824	$	&	$	+18$ $ 42 $ $11.8846	$	&	$	328.80	\pm	2.36	$	&	0.933	&	10.75	&	0.4128220	&$	0.045	\pm	0.001	$	\\
SW Lac	&	$	22$ $ 53 $ $41.6540	$	&	$	+37$ $ 56 $ $18.6075	$	&	$	71.16	\pm	0.10	$	&	1.156	&	8.51	&	0.3207152	&$	0.088	\pm	0.001	$	\\
UV Lyn	&	$	09$ $ 03 $ $24.1259	$	&	$	+38$ $ 05 $ $54.5972	$	&	$	143.65	\pm	0.36	$	&	0.877	&	9.41	&	0.4149809	&$	0.036	\pm	0.001	$	\\
UZ Leo	&	$	10$ $ 40 $ $33.1850	$	&	$	+13$ $ 34 $ $00.8550	$	&	$	329.21	\pm	2.30	$	&	1.024	&	9.58	&	0.6180570	&$	0.089	\pm	0.001	$	\\
V2364 Cyg	&	$	19$ $ 22 $ $11.7423	$	&	$	+49$ $ 28 $ $34.3833	$	&	$	690.83	\pm	6.18	$	&	0.870	&	11.20	&	0.5921376	&$	0.168	\pm	0.001	$	\\
V369 Cep	&	$	00$ $ 46 $ $12.1209	$	&	$	+85$ $ 14 $ $01.9788	$	&	$	1864.87	\pm	107.73	$	&	0.894	&	16.00	&	0.3281916	&$	0.244	\pm	0.001	$	\\
V402 Aur	&	$	05$ $ 02 $ $14.7413	$	&	$	+31$ $ 15 $ $49.2928	$	&	$	192.36	\pm	0.69	$	&	1.012	&	8.84	&	0.6034990	&$	0.307	\pm	0.001	$	\\
V404 Peg	&	$	22$ $ 56 $ $30.8893	$	&	$	+33$ $ 55 $ $12.0956	$	&	$	265.33	\pm	1.79	$	&	1.107	&	10.47	&	0.4191950	&$	0.163	\pm	0.001	$	\\
V502 Oph	&	$	16$ $ 41 $ $20.8609	$	&	$	+00$ $ 30 $ $27.3843	$	&	$	96.39	\pm	0.19	$	&	0.845	&	8.34	&	0.4533880	&$	0.092	\pm	0.001	$	\\
V530 And	&	$	01$ $ 27 $ $41.0685	$	&	$	+33$ $ 51 $ $55.2302	$	&	$	956.87	\pm	12.98	$	&	1.079	&	12.60	&	0.5771072	&$	0.140	\pm	0.001	$	\\
V604 Car	&	$	07$ $ 14 $ $50.6318	$	&	$	-59$ $ 16 $ $04.1298	$	&	$	168.96	\pm	0.31	$	&	0.981	&	9.16	&	0.4722920	&$	0.154	\pm	0.001	$	\\
V842 Her	&	$	16$ $ 06 $ $02.2124	$	&	$	+50$ $ 11 $ $13.1044	$	&	$	171.43	\pm	0.41	$	&	1.077	&	9.85	&	0.4190370	&$	0.037	\pm	0.001	$	\\
VZ Tri	&	$	02$ $ 21 $ $29.8550	$	&	$	+31$ $ 58 $ $46.1741	$	&	$	520.54	\pm	3.76	$	&	1.048	&	12.80	&	0.4334818	&$	0.176	\pm	0.001	$	\\
XY LMi	&	$	10$ $ 34 $ $12.3398	$	&	$	+32$ $ 08 $ $51.6880	$	&	$	385.24	\pm	2.59	$	&	0.902	&	10.71	&	0.4368897	&$	0.043	\pm	0.001	$	\\
\hline
\hline
\end{tabular}
\end{center}
\label{tab1}
\end{table*}


\begin{table*}
\caption{The reference of each of the selected systems and studies.}
\centering
\begin{center}
\footnotesize
\begin{tabular}{c c | c c | c c}
\hline
\hline
System & Reference & System & Reference & System & Reference\\
\hline
AK Her	&	\cite{2014AJ....148..126C}	&	EI CVn	&	\cite{2011RAA....11..181Y}	&	RV CVn	&	\cite{2014ASPC..482..163L}	\\
AO Aqr	&	\cite{2015NewA...41....1U}	&	EQ Cep	&	\cite{2016AJ....152..129C}	&	RV Psc	&	\cite{2009ApSS.321..209H}	\\
AP Leo	&	\cite{2003AA...412..465K}	&	GSC 3553-845	&	\cite{2014PASJ...66..100G}	&	SS Com	&	\cite{2006AJ....131.1032Q}	\\
AQ Tuc	&	\cite{2001CoSka..31....5C}	&	GSC 804-118	&	\cite{2005ApSS.300..337Y}	&	SW Lac	&	\cite{2005AcA....55..123G}	\\
ASAS J035020-8017.4	&	\cite{2016NewA...46...94S}	&	GSC 2936-478	&	\cite{2005ApSS.300..337Y}	&	UV Lyn	&	\cite{2005AcA....55..389Z}	\\
ASAS J063546+1928.6	&	\cite{2018AJ....155..172S}	&	GW Cnc	&	\cite{2016NewA...46...31G}	&	UZ Leo	&	\cite{2018PASP..130c4201L}	\\
AU Ser	&	\cite{2018IBVS.6256....1A}	&	GZ And	&	\cite{2004AcA....54..195B}	&	V2364 Cyg	&	\cite{2002IBVS.5285....1N}	\\
BO Ari	&	\cite{2021NewA...8601571P}	&	HH UMa	&	\cite{2015NewA...34..271Y}	&	V369 Cep	&	\cite{2016AJ....152..129C}	\\
BX Dra	&	\cite{2013PASJ...65....1P}	&	HI Dra	&	\cite{2014AJ....148..126C}	&	V402 Aur	&	\cite{2004AcA....54..299Z}	\\
CU Tau	&	\cite{2005AJ....130..224Q}	&	HI Pup	&	\cite{2014NewA...31...56U}	&	V404 Peg	&	\cite{2011AN....332..690G}	\\
DD Com	&	\cite{2010AJ....140..215Z}	&	KIC 3221207	&	\cite{2017AIPC.1815h0004A}	&	V502 Oph	&	\cite{2016PASJ...68..102X}	\\
DN Boo	&	\cite{2008NewA...13..468S}	&	MQ UMa	&	\cite{2015AJ....150...83Z}	&	V530 And	&	\cite{2016JAVSO..44..108S}	\\
DN Cam	&	\cite{2004AcA....54..195B}	&	NO Cam	&	\cite{2017PASJ...69...37Z}	&	V604 Car	&	\cite{2006ApSS.301..195W}	\\
EF Boo	&	\cite{2005AcA....55..123G}	&	NR Cam	&	\cite{2015NewA...37...64T}	&	V842 Her	&	\cite{2009NewA...14..321E}	\\
EF Dra	&	\cite{2012RAA....12..419Y}	&	OO Aql	&	\cite{2016RAA....16....2L}	&	VZ Tri	&	\cite{2010ApSS.326..125Y}	\\
EH Cnc	&	\cite{2011PASP..123..895Y}	&	RR Cen	&	\cite{2005PASJ...57..983Y}	&	XY LMi	&	\cite{2011AJ....141..151Q}	\\
\hline
\hline
\end{tabular}
\end{center}
\label{tab2}
\end{table*}


\begin{table*}
\caption{Light curves solutions in selected studies.}
\centering
\begin{center}
\footnotesize
\begin{tabular}{c | c | c | c | c | c | c | c | c | c | c}
\hline
\hline
System & $T_1(K)$ & $T_2(K)$ & $q=M_2/M_1$ & $r_{mean1}$ & $r_{mean2}$ & $l_1/l_{tot}$ & $i^{\circ}$ & $f$ & $\Omega$ & $l_3$\\
\hline
AK Her	&	6500	&$	6180	\pm	10	$&$	0.277	\pm	0.024	$&$	0.475			$&$	0.269			$&$	0.782			$&	81.70	&	0.332	&	2.357	&	No	\\
AO Aqr	&	5750	&$	5708	\pm	33	$&$	0.288	\pm	0.006	$&$	0.530	\pm	0.006	$&$	0.325	\pm	0.020	$&$	0.753	\pm	0.007	$&	79.40	&		&	2.310	&	No	\\
AP Leo	&	6150	&$	6250	\pm	25	$&$	0.297			$&$	0.510	\pm	0.001	$&$	0.293	\pm	0.001	$&$	0.746	\pm	0.001	$&	78.00	&	0.060	&	2.434	&	Yes	\\
AQ Tuc	&	6900	&$	7048	\pm	13	$&$	0.354			$&$	0.494	\pm	0.001	$&$	0.316	\pm	0.001	$&$	0.697	\pm	0.001	$&	75.89	&	0.366	&	2.501	&	No	\\
ASAS J035020-8017.4	&	5790	&$	5675	\pm	42	$&$	0.232	\pm	0.005	$&$	0.526	\pm	0.008	$&$	0.278	\pm	0.017	$&$	0.790	\pm	0.010	$&	78.60	&	0.340	&	2.260	&	No	\\
ASAS J063546+1928.6	&	6350	&$	6139	\pm	36	$&$	0.153	\pm	0.002	$&$	0.558	\pm	0.003	$&$	0.267	\pm	0.007	$&$	0.834	\pm	0.057	$&	79.24	&	0.574	&	2.100	&	Yes	\\
AU Ser	&	5140	&$	4986	\pm	1	$&$	0.692	\pm	0.006	$&$	0.412	\pm	0.001	$&$	0.347	\pm	0.001	$&$	0.623	\pm	0.001	$&	82.43	&	0.040	&	3.213	&	No	\\
BO Ari	&	5873	&$	5850	\pm	35	$&$	0.207	\pm	0.001	$&$	0.553	\pm	0.002	$&$	0.295	\pm	0.004	$&$	0.782	\pm	0.001	$&	82.18	&	0.757	&	2.151	&	No	\\
BX Dra	&	6980	&$	6979	\pm	2	$&$	0.288	\pm	0.001	$&$	0.517	\pm	0.001	$&$	0.308	\pm	0.001	$&$	0.744	\pm	0.001	$&	80.63	&		&	2.348	&	No	\\
CU Tau	&	5900	&$	5938	\pm	10	$&$	0.177	\pm	0.002	$&$	0.553	\pm	0.002	$&$	0.265	\pm	0.005	$&$	0.810	\pm	0.001	$&	73.95	&	0.501	&	2.118	&	No	\\
DD Com	&	5500	&$	4995	\pm	31	$&$	3.690	\pm	0.010	$&$	0.276	\pm	0.002	$&$	0.500	\pm	0.002	$&$	0.325	\pm	0.004	$&	47.90	&	0.088	&	7.460	&	No	\\
DN Boo	&	6095	&$	6071	\pm	52	$&$	0.103			$&$	0.599	\pm	0.020	$&$	0.235	\pm	0.040	$&$	0.875	\pm	0.041	$&	60.02	&		&	1.925	&	No	\\
DN Cam	&	6700	&$	6530	\pm	23	$&$	2.260			$&$	0.345	\pm	0.004	$&$	0.484	\pm	0.003	$&$	0.353	\pm	0.002	$&	73.10	&	0.330	&	5.319	&	No	\\
EF Boo	&	6450	&$	6425	\pm	14	$&$	1.871			$&$	0.356	\pm	0.001	$&$	0.435	\pm	0.001	$&$	0.367	\pm	0.002	$&	75.70	&	0.180	&	4.921	&	No	\\
EF Dra	&	6250	&$	6186	\pm	7	$&$	0.160			$&$	0.559	\pm	0.001	$&$	0.255	\pm	0.001	$&$	0.836	\pm	0.009	$&	77.80	&	0.467	&	2.082	&	Yes	\\
EH Cnc	&	6820	&$	6666	\pm	5	$&$	2.510	\pm	0.020	$&$	0.319	\pm	0.001	$&$	0.478	\pm	0.001	$&$	0.327	$&	81.55	&	0.277	&	5.790	&	No	\\
EI CVn	&	4410	&$	4341	\pm	6	$&$	0.461	\pm	0.003	$&$	0.461	\pm	0.003	$&$	0.327	\pm	0.003	$&$	0.639	\pm	0.002	$&	84.50	&	0.210	&	2.742	&	No	\\
EQ Cep	&	5275	&$	4975	\pm	9	$&$	2.090	\pm	0.01	$&$	0.320	\pm	0.002	$&$	0.449	\pm	0.002	$&$	0.410	\pm	0.010	$&	81.40	&		&	5.350	&	No	\\
GSC 3553-845	&	6250	&$	6044	\pm	10	$&$	2.904	\pm	0.012	$&$	0.308	\pm	0.003	$&$	0.491	\pm	0.002	$&$	0.312	\pm	0.001	$&	73.14	&	0.295	&	6.306	&	No	\\
GSC 804-118	&	5800	&$	5964	\pm	24	$&$	0.243	\pm	0.003	$&$	0.515	\pm	0.003	$&$	0.274	\pm	0.006	$&$	0.758	\pm	0.007	$&	84.85	&	0.200	&	2.306	&	No	\\
GSC 2936-478	&	6400	&$	6515	\pm	58	$&$	0.395	\pm	0.011	$&$	0.468	\pm	0.006	$&$	0.307	\pm	0.009	$&$	0.665	\pm	0.008	$&	67.60	&	0.116	&	2.640	&	No	\\
GW Cnc	&	5790	&$	5649	\pm	6	$&$	3.773	\pm	0.007	$&$	0.281	\pm	0.001	$&$	0.508	\pm	0.001	$&$	0.245	\pm	0.001	$&	83.38	&	0.094	&	7.484	&	No	\\
GZ And	&	6200	&$	5810	\pm	25	$&$	1.880			$&$	0.332	\pm	0.002	$&$	0.443	\pm	0.001	$&$	0.428	\pm	0.003	$&	87.00	&	0.080	&	5.011	&	No	\\
HH UMa	&	6550	&$	6297	\pm	40	$&$	0.295			$&$	0.470	\pm	0.044	$&$	0.304	\pm	0.001	$&$	0.740	\pm	0.040	$&	52.40	&	0.450	&	2.379	&	Yes	\\
HI Dra	&	7000	&$	6550	\pm	20	$&$	0.250	\pm	0.005	$&$	0.478			$&$	0.257			$&$	0.818			$&	54.74	&	0.230	&	2.317	&	No	\\
HI Pup	&	6500	&$	6377	\pm	24	$&$	0.206	\pm	0.001	$&$	0.528	\pm	0.010	$&$	0.262	\pm	0.012	$&$	0.816	\pm	0.049	$&	82.20	&		&	2.221	&	No	\\
KIC 3221207	&	6414	&$	6404	\pm	5	$&$	0.244	\pm	0.001	$&$	0.541	\pm	0.004	$&$	0.304	\pm	0.001	$&$	0.763	\pm	0.007	$&	80.85	&	0.790	&	2.218	&	No	\\
MQ UMa	&	6352	&$	6116	\pm	12	$&$	0.211	\pm	0.006	$&$	0.536	\pm	0.003	$&$	0.274	\pm	0.019	$&$	0.821	\pm	0.001	$&	60.70	&	0.413	&	2.204	&	Yes	\\
NO Cam	&	6530	&$	6486	\pm	3	$&$	0.439	\pm	0.001	$&$	0.490	\pm	0.001	$&$	0.350	\pm	0.002	$&$	0.673	\pm	0.001	$&	84.50	&	0.555	&	2.610	&	No	\\
NR Cam	&	5180	&$	5750	\pm	90	$&$	1.062	\pm	0.001	$&$	0.383			$&$	0.373			$&$	0.486	\pm	0.006	$&	66.13	&	0.006	&	3.847	&	No	\\
OO Aql	&	6100	&$	5926	\pm	4	$&$	0.846\pm0.007 $&$	0.420	\pm	0.001	$&$	0.391	\pm	0.001	$&$	0.569	\pm	0.001	$&	87.72	&	0.370	&	3.348	&	No	\\
RR Cen	&	6912	&$	6891	\pm	13	$&$	0.205	\pm	0.004	$&$	0.535	\pm	0.003	$&$	0.268	\pm	0.009	$&$	0.803	\pm	0.019	$&	81.00	&	0.351	&	2.199	&	No	\\
RV CVn	&	4750	&$	4607	\pm	7	$&$	1.740	\pm	0.005	$&$	0.337	\pm	0.002	$&$	0.435	\pm	0.001	$&$	0.413	\pm	0.002	$&	86.40	&	0.098	&	4.822	&	No	\\
RV Psc	&	6300	&$	6283	\pm	23	$&$	0.598	\pm	0.010	$&$	0.428	\pm	0.005	$&$	0.338	\pm	0.007	$&$	0.618	\pm	0.002	$&	84.20	&	0.058	&	3.039	&	No	\\
SS Com	&	6750	&$	6699	\pm	17	$&$	0.286	\pm	0.002	$&$	0.517	\pm	0.002	$&$	0.305	\pm	0.003	$&$	0.752	\pm	0.001	$&	83.66	&	0.496	&	2.346	&	No	\\
SW Lac	&	5800	&$	5515	\pm	13	$&$	1.270			$&$	0.388	\pm	0.001	$&$	0.430	\pm	0.001	$&$	0.504	\pm	0.001	$&	79.80	&	0.300	&	3.977	&	No	\\
UV Lyn	&	6000	&$	5770	\pm	5	$&$	2.685			$&$	0.299	\pm	0.001	$&$	0.480	\pm	0.001	$&$	0.328	\pm	0.001	$&	67.60	&	0.180	&	6.080	&	No	\\
UZ Leo	&	6980	&$	6772	\pm	250	$&$	0.306	\pm	0.005	$&$	0.528	\pm	0.001	$&$	0.334	\pm	0.002	$&$	0.678	\pm	0.002	$&	87.35	&	0.759	&	2.335	&	Yes	\\
V2364 Cyg	&	6820	&$	6615	\pm	20	$&$	0.306	\pm	0.002	$&$	0.510	\pm	0.002	$&$	0.309	\pm	0.004	$&$	0.753	\pm	0.001	$&	81.80	&		&	2.390	&	No	\\
V369 Cep	&	5546	&$	5088	\pm	12	$&$	1.900	\pm	0.01	$&$	0.334	\pm	0.003	$&$	0.446	\pm	0.003	$&$	0.470	\pm	0.020	$&	74.71	&		&	5.010	&	No	\\
V402 Aur	&	6700	&$	6775	\pm	31	$&$	5.008			$&$	0.251	\pm	0.001	$&$	0.524	\pm	0.001	$&$	0.182	\pm	0.028	$&	52.65	&	0.030	&	9.154	&	No	\\
V404 Peg	&	6340	&$	6154	\pm	7	$&$	0.243	\pm	0.008	$&$	0.520	\pm	0.001	$&$	0.281	\pm	0.001	$&$	0.748	\pm	0.009	$&	62.21	&	0.321	&	2.288	&	No	\\
V502 Oph	&	6140	&$	5922	\pm	6	$&$	2.985			$&$	0.303	\pm	0.001	$&$	0.491	\pm	0.001	$&$	0.309	\pm	0.001	$&	72.80	&	0.255	&	6.438	&	No	\\
V530 And	&	6750	&$	6030	\pm	30	$&$	0.386	\pm	0.004	$&$	0.466	\pm	0.001	$&$	0.301	\pm	0.003	$&$	0.792	\pm	0.010	$&	86.70	&	0.050	&	2.637	&	Yes	\\
V604 Car	&	6383	&$	6339	\pm	22	$&$	0.220	\pm	0.003	$&$	0.534			$&$	0.280			$&$	0.798			$&	81.80	&		&	2.229	&	No	\\
V842 Her	&	6020	&$	5723	\pm	10	$&$	3.859	\pm	0.003	$&$	0.282	\pm	0.001	$&$	0.512	\pm	0.001	$&$	0.255	\pm	0.001	$&	78.30	&	0.250	&	7.572	&	Yes	\\
VZ Tri	&	6240	&$	6345	\pm	18	$&$	0.350	\pm	0.004	$&$	0.489	\pm	0.002	$&$	0.308	\pm	0.004	$&$	0.706	\pm	0.002	$&	80.40	&	0.279	&	2.514	&	No	\\
XY LMi	&	6144	&$	6093	\pm	6	$&$	0.148	\pm	0.001	$&$	0.575	\pm	0.001	$&$	0.265	\pm	0.004	$&$	0.838	\pm	0.001	$&	81.04	&	0.741	&	2.027	&	No	\\
\hline
\hline
\end{tabular}
\end{center}
\label{tab3}
\end{table*}


\begin{table*}
\caption{The absolute parameters calculated in studies of target systems with different methods. Some of the selected systems did not have the estimation of absolute parameters in the reference study, which is not included in this table.}
\centering
\begin{center}
\footnotesize
\begin{tabular}{c | c | c | c | c | c | c | c}
\hline
\hline
System & $M_1(M_{\odot})$ & $M_2(M_{\odot})$ & $R_1(R_{\odot})$ & $R_2(R_{\odot})$ & $L_1(L_{\odot})$ & $L_2(L_{\odot})$ & $a(R_{\odot})$ \\
\hline
AO Aqr	&$	1.190			$&$	0.340	\pm	0.010	$&$	1.640	\pm	0.020	$&$	0.900	\pm	0.010	$&$	2.620	\pm	0.060	$&$	0.700	\pm	0.020	$&$	3.100	\pm	0.100	$	\\
AP Leo	&$	1.716	\pm	0.055	$&$	0.899	\pm	0.029	$&$	1.762	\pm	0.036	$&$	1.219	\pm	0.027	$&$				$&$				$&$				$	\\
ASAS J035020-8017.4	&$	0.990			$&$	0.230	\pm	0.010	$&$	1.760	\pm	0.030	$&$	0.870	\pm	0.090	$&$	3.140	\pm	0.090	$&$	0.700	\pm	0.200	$&$				$	\\
AU Ser	&$	0.850	\pm	0.030	$&$	0.590	\pm	0.020	$&$	1.040	\pm	0.010	$&$	0.880	\pm	0.010	$&$	0.675	\pm	0.013	$&$	0.427	\pm	0.009	$&$	2.520	\pm	0.030	$	\\
BO Ari	&$	1.095			$&$	0.227	\pm	0.015	$&$	1.190	\pm	0.007	$&$	0.636	\pm	0.009	$&$	1.517	\pm	0.015	$&$	0.425	\pm	0.011	$&$	2.152	\pm	0.018	$	\\
BX Dra	&$	2.080	\pm	0.100	$&$	0.600	\pm	0.040	$&$	2.130	\pm	0.050	$&$	1.280	\pm	0.030	$&$	9.660	\pm	1.180	$&$	3.050	\pm	0.380	$&$	4.058	\pm	0.087	$	\\
CU Tau	&$	1.200	\pm	0.090	$&$	0.210	\pm	0.020	$&$				$&$				$&$				$&$				$&$				$	\\
DN Boo	&$	1.428	\pm	0.039	$&$	0.148	\pm	0.006	$&$	1.710	\pm	0.067	$&$	0.670	\pm	0.110	$&$	3.750	\pm	0.280	$&$	0.560	\pm	0.170	$&$	2.863	\pm	0.018	$	\\
DN Cam	&$	0.818	\pm	0.015	$&$	1.849	\pm	0.021	$&$	1.224	\pm	0.013	$&$	1.775	\pm	0.016	$&$				$&$				$&$				$	\\
EF Boo	&$	1.547	\pm	0.035	$&$	0.792	\pm	0.026	$&$	1.897	\pm	0.018	$&$	0.837	\pm	0.009	$&$	4.729	\pm	0.090	$&$	1.090	\pm	0.058	$&$				$	\\
EF Dra	&$	1.815	\pm	0.032	$&$	0.290	\pm	0.026	$&$	1.702	\pm	0.002	$&$	0.777	\pm	0.002	$&$	3.961	\pm	0.008	$&$	0.793	\pm	0.004	$&$	3.045	\pm	0.028	$	\\
EQ Cep	&$	0.900	\pm	0.030	$&$	0.430	\pm	0.020	$&$	0.950	\pm	0.060	$&$	0.680	\pm	0.050	$&$	0.620	\pm	0.050	$&$	0.400	\pm	0.090	$&$	2.100	\pm	0.120	$	\\
GW Cnc	&$	0.257	\pm	0.004	$&$	0.971	\pm	0.016	$&$	0.526	\pm	0.003	$&$	0.961	\pm	0.007	$&$	0.279	\pm	0.008	$&$	0.842	\pm	0.010	$&$	1.936	\pm	0.010	$	\\
GZ And	&$	0.593	\pm	0.015	$&$	1.115	\pm	0.018	$&$	0.741	\pm	0.007	$&$	1.005	\pm	0.009	$&$				$&$				$&$				$	\\
HH UMa	&$	1.250	\pm	0.010	$&$	0.370	\pm	0.020	$&$	1.320	\pm	0.010	$&$	0.780	\pm	0.010	$&$	3.030	\pm	0.030	$&$	0.860	\pm	0.040	$&$				$	\\
HI Dra	&$	1.700	\pm	0.300	$&$	0.420	\pm	0.070	$&$	1.970	\pm	0.090	$&$	1.070	\pm	0.050	$&$	7.870	\pm	0.040	$&$	1.800	\pm	0.050	$&$	3.800	\pm	0.200	$	\\
HI Pup	&$	1.210	\pm	0.240	$&$	0.230	\pm	0.190	$&$	1.440	\pm	0.110	$&$	0.670	\pm	0.100	$&$	3.300	\pm	0.500	$&$	0.700	\pm	0.200	$&$	2.700	\pm	0.200	$	\\
KIC 3221207	&$	1.300	\pm	0.100	$&$	0.320	\pm	0.120	$&$	1.670	\pm	0.010	$&$	0.870	\pm	0.020	$&$	4.240	\pm	0.530	$&$	1.150	\pm	0.160	$&$	3.073	\pm	0.025	$	\\
NR Cam	&$	0.920	\pm	0.200	$&$	0.980	\pm	0.200	$&$	0.790	\pm	0.070	$&$	0.810	\pm	0.070	$&$				$&$				$&$				$	\\
OO Aql	&$	1.060	\pm	0.007	$&$	0.897	\pm	0.006	$&$	1.406	\pm	0.002	$&$	1.309	\pm	0.002	$&$	2.453	\pm	0.007	$&$	1.894	\pm	0.006	$&$				$	\\
RR Cen	&$	1.820	\pm	0.260	$&$	0.380	\pm	0.060	$&$	2.100	\pm	0.010	$&$	1.050	\pm	0.030	$&$	8.890			$&$	2.200			$&$	3.920	\pm	0.190	$	\\
SS Com	&$	1.510			$&$	0.430			$&$				$&$				$&$				$&$				$&$				$	\\
SW Lac	&$	1.240	\pm	0.024	$&$	0.964	\pm	0.021	$&$	1.090	\pm	0.007	$&$	0.976	\pm	0.007	$&$	0.971	\pm	0.029	$&$	0.953	\pm	0.014	$&$				$	\\
UV Lyn	&$	0.501	\pm	0.015	$&$	1.344	\pm	0.025	$&$	0.858	\pm	0.007	$&$	1.376	\pm	0.010	$&$	0.840	\pm	0.010	$&$	1.860	\pm	0.010	$&$				$	\\
UZ Leo	&$	2.010	\pm	0.030	$&$	0.620	\pm	0.010	$&$	2.230	\pm	0.010	$&$	1.400	\pm	0.010	$&$	10.600	\pm	1.500	$&$	3.680	\pm	0.550	$&$				$	\\
V369 Cep	&$	0.930	\pm	0.020	$&$	0.490	\pm	0.020	$&$	1.010	\pm	0.080	$&$	0.760	\pm	0.070	$&$	0.770	\pm	0.060	$&$	0.610	\pm	0.080	$&$	2.250	\pm	0.170	$	\\
V402 Aur	&$	1.638	\pm	0.048	$&$	0.327	\pm	0.023	$&$	0.915	\pm	0.008	$&$	1.997	\pm	0.019	$&$				$&$				$&$				$	\\
V404 Peg	&$	1.175	\pm	0.025	$&$	0.286	\pm	0.006	$&$	1.346	\pm	0.010	$&$	0.710	\pm	0.005	$&$	2.623	\pm	0.012	$&$	0.647	\pm	0.003	$&$				$	\\
V502 Oph	&$	0.460	\pm	0.020	$&$	1.370	\pm	0.020	$&$	0.940	\pm	0.010	$&$	1.510	\pm	0.010	$&$	1.130	\pm	0.020	$&$	2.490	\pm	0.030	$&$				$	\\
V842 Her	&$	0.380	\pm	0.010	$&$	1.450	\pm	0.010	$&$	0.810	\pm	0.010	$&$	1.470	\pm	0.010	$&$	0.770	\pm	0.110	$&$	2.080	\pm	0.320	$&$				$	\\
\hline
\hline
\end{tabular}
\end{center}
\label{tab4}
\end{table*}


\begin{table*}
\caption{Estimation of absolute parameters in this study using Gaia DR3 parallax.}
\centering
\begin{center}
\footnotesize
\begin{tabular}{c | c | c | c | c | c | c | c}
\hline
\hline
System & $L_1(L_{\odot})$ & $L_2(L_{\odot})$ & $R_1(R_{\odot})$ & $R_2(R_{\odot})$ & $M_1(M_{\odot})$ & $M_2(M_{\odot})$ & $a(R_{\odot})$\\
\hline
AK Her	&$	3.503	\pm	0.292	$&$	1.003	\pm	0.081	$&$	1.479	\pm	0.062	$&$	0.875	\pm	0.038	$&$	1.908	\pm	0.227	$&$	0.529	\pm	0.063	$&$	3.184	\pm	0.127	$	\\
AO Aqr	&$	3.333	\pm	0.061	$&$	1.102	\pm	0.032	$&$	1.843	\pm	0.018	$&$	1.076	\pm	0.028	$&$	1.700	\pm	0.094	$&$	0.490	\pm	0.027	$&$	3.394	\pm	0.062	$	\\
AP Leo	&$	2.544	\pm	0.033	$&$	0.857	\pm	0.012	$&$	1.408	\pm	0.010	$&$	0.791	\pm	0.012	$&$	1.136	\pm	0.027	$&$	0.338	\pm	0.008	$&$	2.730	\pm	0.022	$	\\
AQ Tuc	&$	8.039	\pm	0.141	$&$	3.483	\pm	0.056	$&$	1.988	\pm	0.018	$&$	1.254	\pm	0.015	$&$	1.787	\pm	0.042	$&$	0.632	\pm	0.015	$&$	3.996	\pm	0.031	$	\\
ASAS J035020-8017.4	&$	11.695	\pm	0.194	$&$	3.171	\pm	0.103	$&$	3.406	\pm	0.030	$&$	1.846	\pm	0.057	$&$	7.924	\pm	0.443	$&$	1.838	\pm	0.103	$&$	6.557	\pm	0.122	$	\\
ASAS J063546+1928.6	&$	6.391	\pm	0.299	$&$	1.298	\pm	0.025	$&$	2.093	\pm	0.048	$&$	1.009	\pm	0.021	$&$	2.746	\pm	0.136	$&$	0.420	\pm	0.021	$&$	3.765	\pm	0.062	$	\\
AU Ser	&$	0.869	\pm	0.012	$&$	0.558	\pm	0.007	$&$	1.178	\pm	0.009	$&$	1.003	\pm	0.006	$&$	1.261	\pm	0.015	$&$	0.873	\pm	0.011	$&$	2.875	\pm	0.012	$	\\
BO Ari	&$	1.754	\pm	0.027	$&$	0.490	\pm	0.008	$&$	1.282	\pm	0.010	$&$	0.683	\pm	0.013	$&$	1.364	\pm	0.023	$&$	0.282	\pm	0.004	$&$	2.316	\pm	0.013	$	\\
BX Dra	&$	9.003	\pm	0.241	$&$	3.099	\pm	0.075	$&$	2.056	\pm	0.028	$&$	1.207	\pm	0.015	$&$	1.911	\pm	0.061	$&$	0.551	\pm	0.017	$&$	3.948	\pm	0.042	$	\\
CU Tau	&$	2.744	\pm	0.071	$&$	0.640	\pm	0.015	$&$	1.589	\pm	0.021	$&$	0.757	\pm	0.011	$&$	1.577	\pm	0.012	$&$	0.279	\pm	0.002	$&$	2.865	\pm	0.007	$	\\
DD Com	&$	0.266	\pm	0.009	$&$	0.647	\pm	0.036	$&$	0.569	\pm	0.009	$&$	1.076	\pm	0.043	$&$	0.369	\pm	0.026	$&$	1.362	\pm	0.094	$&$	2.107	\pm	0.048	$	\\
DN Boo	&$	5.786	\pm	0.101	$&$	0.829	\pm	0.022	$&$	2.162	\pm	0.018	$&$	0.825	\pm	0.025	$&$	2.739	\pm	0.777	$&$	0.282	\pm	0.080	$&$	3.560	\pm	0.327	$	\\
DN Cam	&$	3.941	\pm	0.042	$&$	7.298	\pm	0.124	$&$	1.476	\pm	0.008	$&$	2.115	\pm	0.033	$&$	1.340	\pm	0.009	$&$	3.027	\pm	0.020	$&$	4.324	\pm	0.009	$	\\
EF Boo	&$	1.549	\pm	0.014	$&$	2.677	\pm	0.036	$&$	0.999	\pm	0.004	$&$	1.323	\pm	0.014	$&$	0.660	\pm	0.011	$&$	1.234	\pm	0.020	$&$	2.923	\pm	0.016	$	\\
EF Dra	&$	4.070	\pm	0.030	$&$	0.803	\pm	0.010	$&$	1.724	\pm	0.007	$&$	0.782	\pm	0.007	$&$	1.872	\pm	0.018	$&$	0.299	\pm	0.003	$&$	3.076	\pm	0.010	$	\\
EH Cnc	&$	1.354	\pm	0.036	$&$	2.808	\pm	0.079	$&$	0.835	\pm	0.012	$&$	1.259	\pm	0.019	$&$	0.396	\pm	0.014	$&$	0.994	\pm	0.035	$&$	2.626	\pm	0.031	$	\\
EI CVn	&$	0.398	\pm	0.005	$&$	0.237	\pm	0.004	$&$	1.083	\pm	0.007	$&$	0.862	\pm	0.010	$&$	2.090	\pm	0.024	$&$	0.963	\pm	0.011	$&$	2.492	\pm	0.010	$	\\
EQ Cep	&$	0.364	\pm	0.040	$&$	0.583	\pm	0.080	$&$	0.724	\pm	0.040	$&$	1.030	\pm	0.074	$&$	0.545	\pm	0.096	$&$	1.138	\pm	0.202	$&$	2.278	\pm	0.133	$	\\
GSC 3553-845	&$	1.333	\pm	0.036	$&$	3.006	\pm	0.089	$&$	0.987	\pm	0.013	$&$	1.585	\pm	0.028	$&$	0.603	\pm	0.017	$&$	1.750	\pm	0.047	$&$	3.216	\pm	0.029	$	\\
GSC 804-118	&$	1.681	\pm	0.063	$&$	0.524	\pm	0.024	$&$	1.287	\pm	0.024	$&$	0.679	\pm	0.020	$&$	1.586	\pm	0.051	$&$	0.386	\pm	0.013	$&$	2.488	\pm	0.027	$	\\
GSC 2936-478	&$	2.405	\pm	0.060	$&$	1.202	\pm	0.044	$&$	1.264	\pm	0.016	$&$	0.862	\pm	0.032	$&$	1.044	\pm	0.011	$&$	0.412	\pm	0.005	$&$	2.754	\pm	0.010	$	\\
GW Cnc	&$	0.233	\pm	0.004	$&$	0.735	\pm	0.016	$&$	0.481	\pm	0.005	$&$	0.897	\pm	0.012	$&$	0.187	\pm	0.005	$&$	0.704	\pm	0.017	$&$	1.739	\pm	0.014	$	\\
GZ And	&$	0.484	\pm	0.005	$&$	0.678	\pm	0.013	$&$	0.604	\pm	0.004	$&$	0.814	\pm	0.015	$&$	0.306	\pm	0.008	$&$	0.575	\pm	0.014	$&$	1.828	\pm	0.015	$	\\
HH UMa	&$	2.434	\pm	0.086	$&$	0.872	\pm	0.016	$&$	1.214	\pm	0.021	$&$	0.786	\pm	0.017	$&$	1.269	\pm	0.180	$&$	0.374	\pm	0.053	$&$	2.585	\pm	0.121	$	\\
HI Dra	&$	11.717	\pm	0.194	$&$	2.657	\pm	0.036	$&$	2.332	\pm	0.020	$&$	1.268	\pm	0.016	$&$	3.550	\pm	0.083	$&$	0.888	\pm	0.021	$&$	4.906	\pm	0.038	$	\\
HI Pup	&$	3.477	\pm	0.157	$&$	0.791	\pm	0.009	$&$	1.473	\pm	0.033	$&$	0.730	\pm	0.010	$&$	1.288	\pm	0.144	$&$	0.265	\pm	0.030	$&$	2.788	\pm	0.103	$	\\
KIC 3221207	&$	6.003	\pm	0.080	$&$	1.867	\pm	0.020	$&$	1.988	\pm	0.014	$&$	1.112	\pm	0.004	$&$	2.367	\pm	0.024	$&$	0.577	\pm	0.007	$&$	3.666	\pm	0.013	$	\\
MQ UMa	&$	5.321	\pm	0.187	$&$	1.187	\pm	0.038	$&$	1.909	\pm	0.034	$&$	0.972	\pm	0.019	$&$	2.194	\pm	0.125	$&$	0.463	\pm	0.026	$&$	3.554	\pm	0.067	$	\\
NO Cam	&$	2.635	\pm	0.060	$&$	1.284	\pm	0.027	$&$	1.271	\pm	0.015	$&$	0.899	\pm	0.011	$&$	0.864	\pm	0.020	$&$	0.380	\pm	0.009	$&$	2.582	\pm	0.020	$	\\
NR Cam	&$	0.383	\pm	0.001	$&$	0.352	\pm	0.009	$&$	0.770	\pm	0.001	$&$	0.599	\pm	0.027	$&$	0.587	\pm	0.030	$&$	0.624	\pm	0.032	$&$	1.808	\pm	0.031	$	\\
OO Aql	&$	1.434	\pm	0.017	$&$	1.109	\pm	0.013	$&$	1.074	\pm	0.006	$&$	1.001	\pm	0.007	$&$	0.473	\pm	0.007	$&$	0.401	\pm	0.005	$&$	2.558	\pm	0.011	$	\\
RR Cen	&$	10.578	\pm	0.054	$&$	2.597	\pm	0.035	$&$	2.273	\pm	0.005	$&$	1.133	\pm	0.012	$&$	2.310	\pm	0.106	$&$	0.473	\pm	0.022	$&$	4.238	\pm	0.065	$	\\
RV CVn	&$	0.251	\pm	0.006	$&$	0.387	\pm	0.012	$&$	0.741	\pm	0.008	$&$	0.979	\pm	0.018	$&$	0.742	\pm	0.024	$&$	1.291	\pm	0.040	$&$	2.225	\pm	0.023	$	\\
RV Psc	&$	3.458	\pm	0.094	$&$	2.142	\pm	0.063	$&$	1.564	\pm	0.022	$&$	1.238	\pm	0.027	$&$	1.339	\pm	0.008	$&$	0.800	\pm	0.005	$&$	3.658	\pm	0.008	$	\\
SS Com	&$	3.281	\pm	0.075	$&$	1.084	\pm	0.023	$&$	1.327	\pm	0.016	$&$	0.775	\pm	0.013	$&$	1.019	\pm	0.021	$&$	0.292	\pm	0.006	$&$	2.554	\pm	0.018	$	\\
SW Lac	&$	0.917	\pm	0.011	$&$	0.952	\pm	0.014	$&$	0.950	\pm	0.006	$&$	1.071	\pm	0.013	$&$	0.866	\pm	0.018	$&$	1.099	\pm	0.022	$&$	2.470	\pm	0.017	$	\\
UV Lyn	&$	0.994	\pm	0.012	$&$	2.103	\pm	0.030	$&$	0.925	\pm	0.006	$&$	1.454	\pm	0.013	$&$	0.606	\pm	0.009	$&$	1.628	\pm	0.024	$&$	3.061	\pm	0.015	$	\\
UZ Leo	&$	8.970	\pm	0.253	$&$	3.012	\pm	0.094	$&$	2.052	\pm	0.176	$&$	1.263	\pm	0.113	$&$	1.515	\pm	0.387	$&$	0.464	\pm	0.119	$&$	3.834	\pm	0.322	$	\\
V2364 Cyg	&$	10.686	\pm	0.295	$&$	3.542	\pm	0.090	$&$	2.346	\pm	0.033	$&$	1.436	\pm	0.027	$&$	2.897	\pm	0.070	$&$	0.886	\pm	0.022	$&$	4.624	\pm	0.037	$	\\
V369 Cep	&$	0.720	\pm	0.066	$&$	0.927	\pm	0.127	$&$	0.921	\pm	0.042	$&$	1.242	\pm	0.091	$&$	0.914	\pm	0.142	$&$	1.736	\pm	0.271	$&$	2.771	\pm	0.143	$	\\
V402 Aur	&$	2.004	\pm	0.275	$&$	8.978	\pm	0.158	$&$	1.053	\pm	0.072	$&$	2.179	\pm	0.039	$&$	0.447	\pm	0.038	$&$	2.237	\pm	0.190	$&$	4.177	\pm	0.118	$	\\
V404 Peg	&$	3.136	\pm	0.038	$&$	1.075	\pm	0.022	$&$	1.471	\pm	0.009	$&$	0.914	\pm	0.012	$&$	1.727	\pm	0.035	$&$	0.420	\pm	0.009	$&$	3.041	\pm	0.020	$	\\
V502 Oph	&$	1.155	\pm	0.013	$&$	2.650	\pm	0.037	$&$	0.952	\pm	0.006	$&$	1.550	\pm	0.014	$&$	0.511	\pm	0.008	$&$	1.527	\pm	0.022	$&$	3.149	\pm	0.015	$	\\
V530 And	&$	5.808	\pm	0.144	$&$	1.616	\pm	0.058	$&$	1.766	\pm	0.022	$&$	1.167	\pm	0.032	$&$	1.638	\pm	0.070	$&$	0.632	\pm	0.027	$&$	3.834	\pm	0.055	$	\\
V604 Car	&$	4.496	\pm	0.060	$&$	1.141	\pm	0.013	$&$	1.737	\pm	0.012	$&$	0.887	\pm	0.011	$&$	1.630	\pm	0.033	$&$	0.359	\pm	0.008	$&$	3.210	\pm	0.022	$	\\
V842 Her	&$	0.725	\pm	0.008	$&$	2.210	\pm	0.034	$&$	0.784	\pm	0.005	$&$	1.515	\pm	0.017	$&$	0.372	\pm	0.007	$&$	1.434	\pm	0.026	$&$	2.870	\pm	0.017	$	\\
VZ Tri	&$	1.366	\pm	0.029	$&$	0.564	\pm	0.012	$&$	1.002	\pm	0.011	$&$	0.623	\pm	0.011	$&$	0.447	\pm	0.008	$&$	0.156	\pm	0.003	$&$	2.036	\pm	0.011	$	\\
XY LMi	&$	5.425	\pm	0.125	$&$	1.055	\pm	0.019	$&$	2.060	\pm	0.024	$&$	0.924	\pm	0.010	$&$	2.705	\pm	0.023	$&$	0.400	\pm	0.004	$&$	3.535	\pm	0.010	$	\\
\hline
\hline
\end{tabular}
\end{center}
\label{tab5}
\end{table*}


\begin{table*}
\caption{Some other estimated parameters for target systems.}
\centering
\begin{center}
\footnotesize
\begin{tabular}{c | c | c | c | c | c | c | c}
\hline
\hline
System & $a_1(R_{\odot})$ & $a_2(R_{\odot})$ & $\Delta a(R_{\odot})$ & $M_{V(System)}$ & $log(g)_1$ & $log(g)_2$ & $J_0$\\
\hline
AK Her	&	3.114	&	3.253	&	0.139	&$	3.114	\pm	0.072	$&$	4.379	\pm	0.089	$&$	4.278	\pm	0.090	$&$	51.843	\pm	0.108	$	\\
AO Aqr	&	3.477	&	3.311	&	0.166	&$	3.206	\pm	0.010	$&$	4.137	\pm	0.015	$&$	4.065	\pm	0.002	$&$	51.797	\pm	0.045	$	\\
AP Leo	&	2.761	&	2.700	&	0.061	&$	3.438	\pm	0.004	$&$	4.196	\pm	0.017	$&$	4.171	\pm	0.023	$&$	51.499	\pm	0.018	$	\\
AQ Tuc	&	4.024	&	3.968	&	0.056	&$	2.056	\pm	0.000	$&$	4.093	\pm	0.018	$&$	4.042	\pm	0.021	$&$	51.943	\pm	0.018	$	\\
ASAS J035020-8017.4	&	6.475	&	6.640	&	0.165	&$	1.891	\pm	0.012	$&$	4.273	\pm	0.017	$&$	4.170	\pm	0.003	$&$	52.858	\pm	0.046	$	\\
ASAS J063546+1928.6	&	3.751	&	3.779	&	0.028	&$	2.536	\pm	0.003	$&$	4.235	\pm	0.042	$&$	4.054	\pm	0.003	$&$	51.881	\pm	0.040	$	\\
AU Ser	&	2.859	&	2.890	&	0.031	&$	4.631	\pm	0.004	$&$	4.396	\pm	0.012	$&$	4.376	\pm	0.011	$&$	51.888	\pm	0.010	$	\\
BO Ari	&	2.318	&	2.315	&	0.003	&$	3.933	\pm	0.002	$&$	4.357	\pm	0.015	$&$	4.219	\pm	0.024	$&$	51.441	\pm	0.013	$	\\
BX Dra	&	3.977	&	3.919	&	0.058	&$	2.001	\pm	0.010	$&$	4.093	\pm	0.026	$&$	4.016	\pm	0.024	$&$	51.906	\pm	0.024	$	\\
CU Tau	&	2.873	&	2.857	&	0.016	&$	3.475	\pm	0.009	$&$	4.234	\pm	0.014	$&$	4.125	\pm	0.016	$&$	51.519	\pm	0.009	$	\\
DD Com	&	2.062	&	2.152	&	0.090	&$	5.098	\pm	0.027	$&$	4.495	\pm	0.045	$&$	4.509	\pm	0.065	$&$	51.525	\pm	0.050	$	\\
DN Boo	&	3.609	&	3.511	&	0.098	&$	2.721	\pm	0.012	$&$	4.206	\pm	0.126	$&$	4.055	\pm	0.091	$&$	51.705	\pm	0.215	$	\\
DN Cam	&	4.278	&	4.370	&	0.092	&$	2.107	\pm	0.002	$&$	4.227	\pm	0.007	$&$	4.268	\pm	0.016	$&$	52.387	\pm	0.005	$	\\
EF Boo	&	2.806	&	3.041	&	0.235	&$	3.176	\pm	0.004	$&$	4.258	\pm	0.011	$&$	4.286	\pm	0.017	$&$	51.786	\pm	0.012	$	\\
EF Dra	&	3.084	&	3.067	&	0.017	&$	3.035	\pm	0.000	$&$	4.237	\pm	0.008	$&$	4.127	\pm	0.011	$&$	51.606	\pm	0.009	$	\\
EH Cnc	&	2.618	&	2.634	&	0.016	&$	3.172	\pm	0.012	$&$	4.192	\pm	0.027	$&$	4.235	\pm	0.029	$&$	51.515	\pm	0.024	$	\\
EI CVn	&	2.349	&	2.636	&	0.287	&$	5.919	\pm	0.003	$&$	4.689	\pm	0.008	$&$	4.551	\pm	0.011	$&$	52.041	\pm	0.009	$	\\
EQ Cep	&	2.262	&	2.294	&	0.032	&$	5.070	\pm	0.114	$&$	4.455	\pm	0.124	$&$	4.469	\pm	0.138	$&$	51.639	\pm	0.128	$	\\
GSC 3553-845	&	3.205	&	3.228	&	0.023	&$	3.179	\pm	0.012	$&$	4.230	\pm	0.023	$&$	4.281	\pm	0.027	$&$	51.872	\pm	0.019	$	\\
GSC 804-118	&	2.499	&	2.478	&	0.021	&$	3.952	\pm	0.029	$&$	4.419	\pm	0.030	$&$	4.361	\pm	0.040	$&$	51.618	\pm	0.027	$	\\
GSC 2936-478	&	2.701	&	2.808	&	0.107	&$	3.347	\pm	0.020	$&$	4.253	\pm	0.016	$&$	4.182	\pm	0.037	$&$	51.554	\pm	0.013	$	\\
GW Cnc	&	1.712	&	1.766	&	0.054	&$	4.871	\pm	0.003	$&$	4.346	\pm	0.018	$&$	4.380	\pm	0.022	$&$	51.045	\pm	0.017	$	\\
GZ And	&	1.819	&	1.837	&	0.018	&$	4.623	\pm	0.002	$&$	4.362	\pm	0.016	$&$	4.376	\pm	0.026	$&$	51.185	\pm	0.017	$	\\
HH UMa	&	2.583	&	2.586	&	0.003	&$	3.441	\pm	0.000	$&$	4.373	\pm	0.076	$&$	4.220	\pm	0.042	$&$	51.556	\pm	0.104	$	\\
HI Dra	&	4.879	&	4.934	&	0.055	&$	1.823	\pm	0.001	$&$	4.253	\pm	0.017	$&$	4.180	\pm	0.021	$&$	52.302	\pm	0.022	$	\\
HI Pup	&	2.790	&	2.786	&	0.004	&$	3.159	\pm	0.003	$&$	4.212	\pm	0.068	$&$	4.135	\pm	0.036	$&$	51.442	\pm	0.083	$	\\
KIC 3221207	&	3.675	&	3.658	&	0.017	&$	2.503	\pm	0.005	$&$	4.215	\pm	0.002	$&$	4.107	\pm	0.008	$&$	51.965	\pm	0.009	$	\\
MQ UMa	&	3.562	&	3.547	&	0.015	&$	2.718	\pm	0.018	$&$	4.218	\pm	0.009	$&$	4.128	\pm	0.008	$&$	51.851	\pm	0.049	$	\\
NO Cam	&	2.594	&	2.569	&	0.025	&$	3.249	\pm	0.007	$&$	4.166	\pm	0.021	$&$	4.110	\pm	0.020	$&$	51.456	\pm	0.017	$	\\
NR Cam	&	2.010	&	1.606	&	0.404	&$	5.236	\pm	0.006	$&$	4.434	\pm	0.022	$&$	4.678	\pm	0.061	$&$	51.432	\pm	0.037	$	\\
OO Aql	&	2.557	&	2.560	&	0.003	&$	3.765	\pm	0.005	$&$	4.051	\pm	0.011	$&$	4.040	\pm	0.011	$&$	51.292	\pm	0.011	$	\\
RR Cen	&	4.249	&	4.228	&	0.021	&$	1.913	\pm	0.001	$&$	4.088	\pm	0.022	$&$	4.004	\pm	0.011	$&$	51.911	\pm	0.039	$	\\
RV CVn	&	2.199	&	2.251	&	0.052	&$	5.716	\pm	0.008	$&$	4.569	\pm	0.023	$&$	4.567	\pm	0.029	$&$	51.782	\pm	0.022	$	\\
RV Psc	&	3.654	&	3.663	&	0.009	&$	2.882	\pm	0.013	$&$	4.176	\pm	0.015	$&$	4.156	\pm	0.022	$&$	51.928	\pm	0.007	$	\\
SS Com	&	2.567	&	2.541	&	0.026	&$	3.120	\pm	0.007	$&$	4.200	\pm	0.019	$&$	4.125	\pm	0.023	$&$	51.399	\pm	0.017	$	\\
SW Lac	&	2.448	&	2.491	&	0.043	&$	4.161	\pm	0.006	$&$	4.420	\pm	0.014	$&$	4.419	\pm	0.019	$&$	51.809	\pm	0.015	$	\\
UV Lyn	&	3.094	&	3.029	&	0.065	&$	3.588	\pm	0.004	$&$	4.288	\pm	0.012	$&$	4.325	\pm	0.014	$&$	51.844	\pm	0.010	$	\\
UZ Leo	&	3.886	&	3.781	&	0.105	&$	1.904	\pm	0.006	$&$	3.994	\pm	0.184	$&$	3.902	\pm	0.187	$&$	51.772	\pm	0.193	$	\\
V2364 Cyg	&	4.600	&	4.647	&	0.047	&$	1.835	\pm	0.010	$&$	4.159	\pm	0.023	$&$	4.071	\pm	0.026	$&$	52.234	\pm	0.019	$	\\
V369 Cep	&	2.757	&	2.785	&	0.028	&$	4.403	\pm	0.113	$&$	4.470	\pm	0.107	$&$	4.489	\pm	0.131	$&$	51.992	\pm	0.113	$	\\
V402 Aur	&	4.195	&	4.158	&	0.037	&$	2.112	\pm	0.001	$&$	4.044	\pm	0.096	$&$	4.111	\pm	0.022	$&$	51.877	\pm	0.062	$	\\
V404 Peg	&	2.829	&	3.253	&	0.424	&$	3.188	\pm	0.006	$&$	4.340	\pm	0.014	$&$	4.139	\pm	0.019	$&$	51.717	\pm	0.024	$	\\
V502 Oph	&	3.142	&	3.157	&	0.015	&$	3.328	\pm	0.005	$&$	4.189	\pm	0.012	$&$	4.241	\pm	0.014	$&$	51.768	\pm	0.011	$	\\
V530 And	&	3.790	&	3.877	&	0.087	&$	2.556	\pm	0.020	$&$	4.158	\pm	0.030	$&$	4.105	\pm	0.043	$&$	51.910	\pm	0.033	$	\\
V604 Car	&	3.253	&	3.168	&	0.085	&$	2.867	\pm	0.005	$&$	4.171	\pm	0.014	$&$	4.097	\pm	0.020	$&$	51.652	\pm	0.019	$	\\
V842 Her	&	2.780	&	2.959	&	0.179	&$	3.643	\pm	0.004	$&$	4.220	\pm	0.013	$&$	4.234	\pm	0.018	$&$	51.609	\pm	0.013	$	\\
VZ Tri	&	2.049	&	2.023	&	0.026	&$	4.042	\pm	0.007	$&$	4.087	\pm	0.018	$&$	4.042	\pm	0.022	$&$	50.890	\pm	0.014	$	\\
XY LMi	&	3.583	&	3.487	&	0.096	&$	2.738	\pm	0.006	$&$	4.243	\pm	0.014	$&$	4.109	\pm	0.014	$&$	51.844	\pm	0.008	$	\\
\hline
\hline
\end{tabular}
\end{center}
\label{tab6}
\end{table*}

\clearpage

\begin{table*}
\caption{This study's photometric solution and estimated absolute parameters of the OO Aql contact binary system.}
\centering
\begin{center}
\footnotesize
\begin{tabular}{c c c c c}
 \hline
 \hline
Parameter & Result & & Parameter & Result\\
\hline
$T_{1}$ (K) 	&	$5398\pm23$	&&	$r_{1(mean)}$ 	&	$0.410\pm0.004$	\\
$T_{2}$ (K) 	&	$5445\pm25$	&&	$r_{2(mean)}$ 	&	$0.379\pm0.004$	\\
$q=M_2/M_1$ 	&	$0.842\pm0.022$	&&	$g_1=g_2$	&	0.32	\\
$\Omega_1=\Omega_2$ 	&	$3.415\pm0.019$	&&	$A_1=A_2$	&	0.50	\\
$i^{\circ}$ 	&	$84.81\pm0.65$	&&	$Col.(deg)_{spot}$ 	&	$86\pm1$	\\
$f$ 	&	$0.157\pm0.042$	&&	$Long.(deg)_{spot}$ 	&	$346\pm1$	\\
$l_1/l_{tot}$ 	&	$0.527\pm0.015$	&&	$Rad.(deg)_{spot}$ 	&	$11\pm1$	\\
$l_2/l_{tot}$ 	&	$0.489\pm0.014$	&&	$T_{spot}/T_{star}$ 	&	$0.80\pm1$	\\
\hline							
$M_1(M_\odot)$	&	$1.146\pm0.021$	&&	$M_{bol1}(mag.)$	&	$4.330\pm0.030$	\\
$M_2(M_\odot)$	&	$0.965\pm0.018$	&&	$M_{bol2}(mag.)$	&	$4.391\pm0.032$	\\
$R_1(R_\odot)$	&	$1.384\pm0.018$	&&	$M_{V(system)}(mag.)$	&	$3.765\pm0.005$	\\
$R_2(R_\odot)$	&	$1.322\pm0.019$	&&	$log(g_1)(cgs)$	&	$4.215\pm0.012$	\\
$L_1(L_\odot)$	&	$1.459\pm0.040$	&&	$log(g_2)(cgs)$	&	$4.180\pm0.012$	\\
$L_2(L_\odot)$	&	$1.379\pm0.040$	&&	$a(R_\odot)$	&	$3.432\pm0.021$	\\
\hline
\hline
\end{tabular}
\end{center}
\label{tab7}
\end{table*}

\bibliography{Ref}{}
\bibliographystyle{aasjournal}

\end{document}